\documentclass[preprint,amsmath,amssymb]{revtex4}

\usepackage{graphicx}
\usepackage{dcolumn}% Align table columns on decimal point
\usepackage{bm}% bold math
\usepackage{ulem}

\begin{document}
\title{Post-processed time-delay interferometry for LISA}

\author{D. A. Shaddock}
\email{Daniel.Shaddock@jpl.nasa.gov}
\affiliation{Jet Propulsion Laboratory, California Institute of
  Technology, Pasadena, CA 91109}
\author{B. Ware}
\email{Brent.Ware@jpl.nasa.gov}
\affiliation{Jet Propulsion Laboratory, California Institute of
  Technology, Pasadena, CA 91109}
\author{R. E. Spero}
\email{Robert.E.Spero@jpl.nasa.gov}
\affiliation{Jet Propulsion Laboratory, California Institute of
  Technology, Pasadena, CA 91109}
\author{M. Vallisneri}
\email{Michele.Vallisneri@jpl.nasa.gov}
\affiliation{Jet Propulsion Laboratory, California Institute of
  Technology, Pasadena, CA 91109}

\begin{abstract}
High-precision interpolation of LISA phase measurements allows
signal reconstruction and formulation of Time-Delay Interferometry
(TDI) combinations to be conducted in post-processing.  The
reconstruction is based on phase measurements made at approximately
10~Hz, at regular intervals independent of the TDI delay times.
Interpolation introduces an error less than $1\times 10^{-8}$ with continuous data segments as short as two seconds in duration. Potential simplifications in the design and operation of LISA are presented.
\end{abstract}

\date{\today}
%\pacs{04.80.Nn, 95.55.Ym, and 07.60.Ly}
\maketitle

\section{Introduction}

The Laser Interferometer Space Antenna (LISA) is a mission to detect
gravitational waves in the frequency band from 0.1~mHz to 1~Hz.  The
LISA constellation consists of three spacecraft flying in a
heliocentric, Earth-trailing orbit, with separations of $L\approx
5\times10^{9}$~m.  Each spacecraft contains two proof masses that are
shielded from external disturbances.  To detect a passing
gravitational wave, the change in separation $\delta L$ of the proof masses in different
spacecraft must be monitored with a precision of $\delta L/L \lesssim
10^{-20} /\sqrt{\rm{Hz}}$ using laser interferometry.  This fractional
length stability is far better than the fractional frequency stability
of the laser source, which is expected to be $\delta \nu/\nu \gtrsim
10^{-14} /\sqrt{\rm{Hz}}$.  Degradation in sensitivity due to laser
frequency noise could be avoided by operating the constellation as a
Michelson interferometer with equal armlengths.  Unfortunately, the
orbital dynamics of the constellation make it impracticable to
equalize the LISA armlengths accurately enough to cancel the excess
frequency noise.  Time delay interferometry (TDI) \cite{Tinto} is a
technique to remove the otherwise overwhelming laser frequency
fluctuations.  TDI cancels laser frequency noise by combining phase
measurements made at different times.  The required timing of the measurements
is set by the light travel times between the LISA spacecraft,
and it must be accurate to 100~ns to meet the laser frequency noise
suppression requirements \cite{TintoPRD03}.

One obvious method to achieve this timing accuracy is to measure the
phase with a 10~MHz sampling frequency. Selecting the nearest-neighbor
samples would then provide the requisite $100$~nsec timing
resolution. This approach, however, would require data to be transmitted between spacecraft or
back to Earth at the rate of approximately $10^9$~bits/s. The current design for TDI is to sample the phase at a
much lower data rate, in the range of 2~to 10~Hz, with 100~nsec
accuracy triggering of the phasemeters
\cite{HellingsPRD01,TintoPRD03}.  This approach poses a number of
technical challenges.  To ensure a timing accuracy of 100~ns, the
absolute lengths of the arms must be known to an
accuracy of 30~m when the measurement is made.  For some TDI combinations, each spacecraft must have knowledge of its nonadjacent arm's length.  Also, the clocks
on different spacecraft must be synchronized at the 100~ns level.
Errors in armlength knowledge or clock synchronization would lead to
an irreversible corruption of the TDI combinations.

An alternative approach is to sample the phase with a low rate at
equally spaced times, and to reconstruct the phase at intermediate
times by interpolation.  Interpolation must be implemented with
exceptional accuracy for effective cancellation of laser frequency
noise by subsequent TDI processing.  Tinto and colleagues
\cite{TintoPRD03} examined one possible method of interpolation and
found that months of uninterrupted data around the time of interest
are needed to achieve the necessary accuracy.  This implies that
months of data would be unusable at the beginning and end of a
measurement, and levies extreme requirements on instrument reliability
and operating duty cycle.  The interpolation technique was deemed infeasible, and the triggered measurement approach was adopted.

In this article we demonstrate that interpolation is feasible and that
it can produce the required accuracy with less than two seconds of
data.  We discuss the significant simplification in the design and
operation of the LISA mission resulting from this change.  The method
is based on fractional delay filtering~\cite{Split}, a mature
technique in digital signal processing.

\section{Interpolation by Fractional-Delay Filtering}

We specify that the interpolation error be less than
$1\times10^{-6}$~cycles$/\sqrt{\rm Hz}$ for frequency components from 1~mHz
to 1~Hz.  This noise level is approximately a factor of 10 below the
phase noise contribution of shot noise.  Below 1~mHz the requirement
is relaxed, as the $1/f^{2}$ proof mass displacement noise dominates
shot noise and a larger interpolation error can be tolerated.
Assuming that the laser frequency noise produces approximately
100~cycles$/\sqrt{\rm Hz}$ at the phasemeter output, interpolation
must have a fractional error of less than $1\times10^{-8}$ for
frequency components in the 1~mHz to 1~Hz range.

We assume that the LISA phase measurements will be recorded with a
$f_s=10$~Hz sampling rate.  The sampling rate must be high enough to
accurately reproduce the phase information in the LISA signal band, and
to avoid adding noise from aliasing of higher-frequency phase noise, at the
$10^{-6}$~cycles$/\sqrt{\rm Hz}$ level.  Moreover, the performance of the interpolation schemes considered below improves with oversampling. Ultimately, the sampling rate will be determined by filtering requirements on the phasemeter
and by the availability and cost of telemetry bandwidth to Earth.

\subsection{Perfect interpolation and fractional-delay filters}

Interpolation is the process of reconstructing the amplitude of a
regularly sampled signal between samples.  Shannon \cite{j:shannon}
proved that a bandlimited signal sampled at a sufficiently high frequency can
be reconstructed perfectly by convolving the discrete time series
with a continuous sine cardinal function ${\rm sinc}(f_s t)=\sin(\pi f_s t) / (\pi f_s t)$.

Sinc interpolation can also be viewed as applying an
acausal finite-impulse-response (FIR) filter to the sampled time
series. The filter kernel (impulse response) is a sampled version of the sinc function. In effect, instead of interpolating the signal we interpolate the filter kernel. As the sinc function is a known analytic function, the filter kernel can be interpolated with arbitrary accuracy simply by time shifting the argument of the sinc. In general, the interpolated signal $s(n-D)$ is the discrete convolution of the original signal $s(n)$ with the shifted kernel:
\begin{equation}
s(n - D) = s(n) \ast h(n - D),
\label{eq:conv}
\end{equation}
where $n$ is the sampling index, 
$D$ is the delay in samples $(-\frac 1 2 \le D < \frac 1 2)$ and $h(n)$ is the filter kernel. For sinc interpolation, $h(n-D)={\rm sinc}(n-D)$. 

\begin{figure}[htb]
\begin{center}
\includegraphics[width=5.0 in]{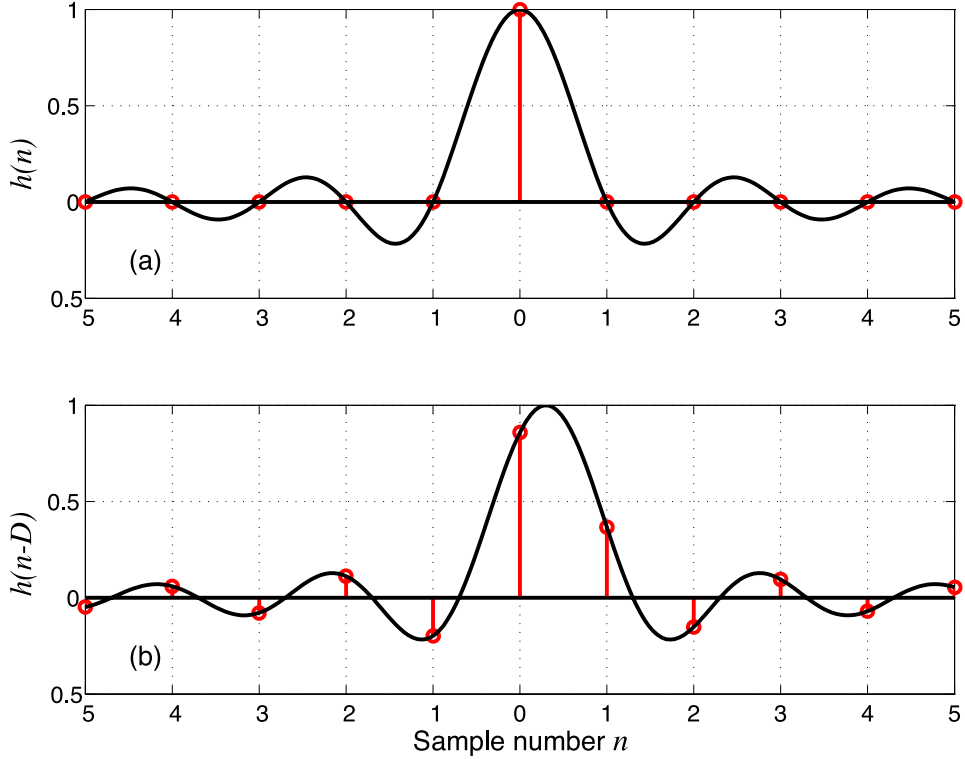}
\caption{Sinc FIR filter kernel values (circles) with delay set to (a)
  $D=0$, and (b) $D=0.3$.}
\label{fig:impulse}
\end{center}
\end{figure}

With zero delay ($D=0$), sinc interpolation corresponds to a FIR filter with
delta function impulse response [see Fig.\ \ref{fig:impulse}(a)],
since for integer $n$ sinc($n$)=$\delta_{n0}$  where $\delta_{nk}$ is the Kronecker delta function.
If $D\neq0$, we obtain a FIR filter
with a non-delta impulse response [see Fig.\ \ref{fig:impulse}(b)],
which has the effect of applying the fractional delay $D$ to the
original time series.  Errors in fractional-delay filtering are caused
by the finite length approximation of the infinitely long
delayed-sinc filter.

\subsection{Truncated-sinc fractional-delay filters}

The simplest finite-length approximation to the ideal delayed-sinc filter is
obtained by truncating the kernel. Filtering by a truncated-sinc of kernel length $N$ can be written as
\begin{equation}
s_N(n - D) = \sum_{k=-\frac{N-1}{2}}^{\frac{N-1}{2}} s(n + k) \,
\textrm{sinc}(D-k) \;\;\;\;\;\;\;\;\; {\rm (for \; odd} \; N).
\label{eq:trunc}
\end{equation}
In the following discussion we restrict ourselves to filters where $N$ is odd for simplicity.

\begin{figure}[tb]
\begin{center}
\includegraphics[width=5.0 in]{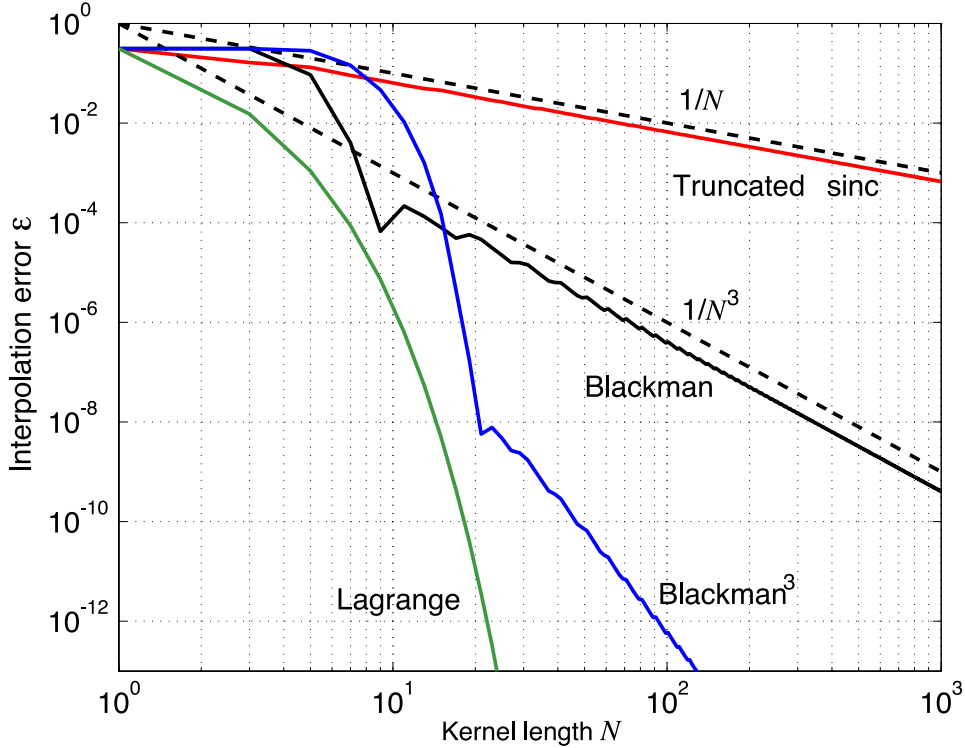}
\caption{\label{fig:herror} Comparison of interpolation error for four
  interpolation methods; truncated-sinc filter kernel, Blackman windowed-sinc filter kernel, Blackman$^3$ windowed-sinc filter kernel, and Lagrange interpolation.}
\end{center}
\end{figure}

Although for a given filter order $N$ the truncated-sinc is optimal in
a least-squares sense~\cite{Split}, its frequency response is far from ideal (unity magnitude), exhibiting
significant ripple even at low frequencies. This is unacceptable for TDI,
where very high fidelity is required in the 1~mHz to 1~Hz measurement
band. In fact, Ref.~\cite{TintoPRD03} showed that truncated-sinc
interpolation becomes sufficiently accurate only for very large $N$.  Figure~\ref{fig:herror} shows the interpolation error versus
$N$ for several filters, including the truncated-sinc.   The interpolation error $\varepsilon$ is defined as the maximum difference of the filter's frequency response and the ideal frequency response, ${\rm e}^{-i2\pi f D/f_s}$ for frequencies between 1~mHz and 1~Hz:
\begin{eqnarray}
\varepsilon={\rm max}(| H(f)-{\rm e}^{-i2\pi f D/f_s} |_{1~{\rm mHz}\leq f \leq 1\; {\rm Hz}})
\end{eqnarray}
where $H(f)$ is the Fourier transform of $h(n)$. We used $D=0.5$, which is expected from theory to be the worst case. With truncated-sinc interpolation, $\varepsilon\approx 1/N$.  Sampling at 10~Hz, this filter would require a kernel almost four months long, $N\gtrsim10^8$, to achieve $\varepsilon < 10^{-8}$.  This means that two months of data at the beginning and end of each measurement period would be unusable.

\subsection{Windowed-sinc fractional-delay filters}
\label{sec:winsinc}

The ripple in the frequency response of the truncated-sinc filter can
be significantly reduced by windowing the filter,
\begin{equation}
s_N(n - D) = \sum_{k=-\frac{N-1}{2}}^{\frac{N-1}{2}} s(n + k) \,
w(k) \, \textrm{sinc}(D-k), \label{eq:winsinc}
\end{equation}
where the window $w(k)$ goes smoothly to zero for $k=\pm(N-1)/2$, so
that the endpoints are tapered to zero instead of abruptly
truncated.  Of several conventional windows~\cite{Smith} tested, the
Blackman function
\begin{eqnarray}
w_b(n)&=&0.42+0.5\cos\left(\frac{\pi n}{N-1}\right) +0.08 \cos \left(\frac{2\pi n}{N-1}\right)
\end{eqnarray}
was best, producing $\varepsilon<10^{-8}$ for $N\geq345$ (see Fig. \ref{fig:herror}) corresponding to a loss of 14.4~seconds of data at the beginning and end of each measurement period.  In
comparison, the TDI combinations need several $L/c$ arm travel times, or at
least 65~seconds, to gather enough data to cancel laser noise.  As seen
in Fig.~\ref{fig:herror}, $\varepsilon \sim 1/N^3$ for Blackman windowed-sinc filters.

One simple modification to the Blackman windowed-sinc filter kernel is to apply the Blackman function more than once. Our tests showed that using $w_b^3(n)$ (applying the Blackman three times) produced $\varepsilon<10^{-8}$ for $N\geq21$, corresponding to a loss of 1.1~seconds of data at the beginning and end of each measurement period.

\subsection{Lagrange filter}

A more accurate filter at low frequencies can be found by requiring a
maximally flat frequency response at DC~\cite{Split}.  This filter kernel is equal to the Lagrange polynomial ~\cite{Koot96}
\begin{equation}
h_L(n) = \prod_{\substack{k = \frac{N-1}{2}\\k \ne n}}^{\frac{N-1}{2}} 
         \frac{t_D - k}{n - k}, \label{eq:lag}
\end{equation}
where $t_D = \frac{N-1}{2} + D$. The Lagrange filter can be compared to filters in Section~\ref{sec:winsinc} by expressing its kernel as a windowed-sinc function (Eq. \ref{eq:winsinc}), with the window
\begin{equation}
w_L(n) = \frac{\pi N}{\sin \pi t_D} {t_D \choose N} {N-1 \choose n + (N-1)/2}, 
\label{eq:lagrange}
\end{equation}
where the binomial coefficient is extended to non-integer
arguments by the generalized factorial function (Gamma function)~\cite{Gamma}.  
\begin{figure}
\begin{center}
\includegraphics[width=5.0in]{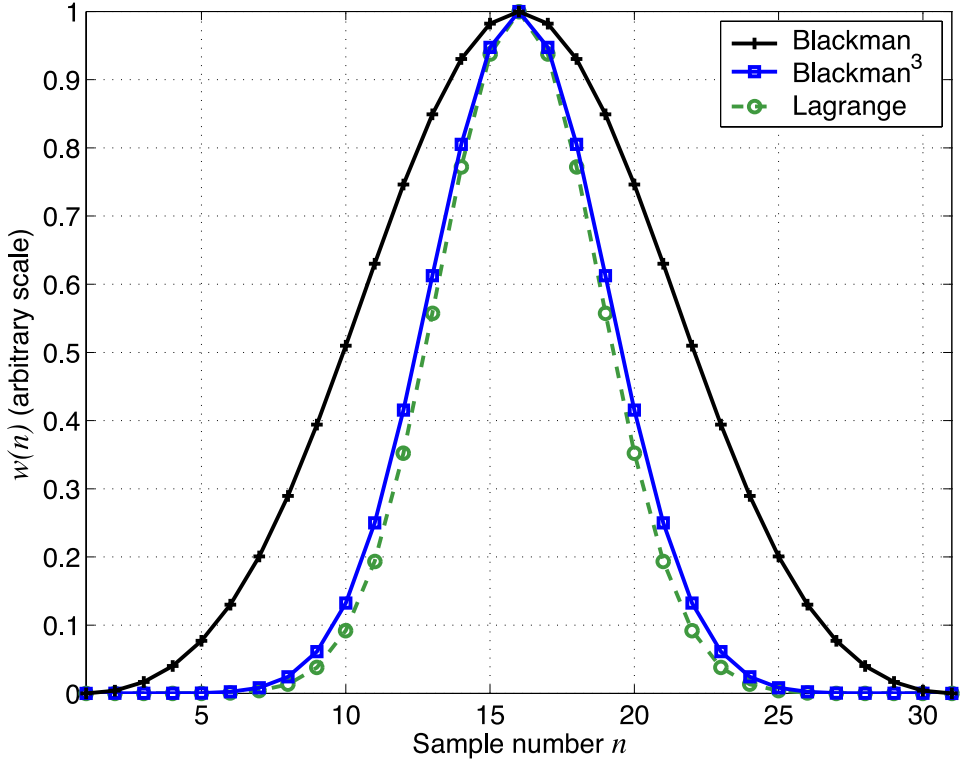}
\caption{\label{fig:windows} Blackman, Blackman$^3$ and Lagrange windows with $N = 31$.}
\end{center}
\end{figure}
Figure~\ref{fig:windows} shows the $w_b(n)$, $w_b^3(n)$ and $w_L(n)$ with $N=31$. 

The performance of the Lagrange filter for this application is excellent, meeting the
requirements with $N = 15$ (1.5 seconds) as shown in Fig. \ref{fig:herror}.  Although Lagrange interpolation is known
to produce large spurious oscillations at the ends of the
interpolation interval, this does not occur when the kernel is fully
immersed in the signal.  By accepting only data where the kernel is
completely immersed, we obtain excellent performance at the expense of
losing $N/2$ points at the beginning and end of each measurement
period.

Lagrange interpolation is related also to the Thiran
infinite-impulse-response (IIR) fractional-delay filter~\cite{Split},
which has nominally flat frequency response. The performance of the
Thiran filter in our application is comparable to the Lagrange
windowed-sinc FIR of the same order; but the latter is favored on grounds of simplicity,
especially for time-dependent delays.

The filters were tested both by interpolating known analytic functions and interpolating bandlimited white noise. The bandlimited noise was generated with a sampling rate of 10~MHz and resampled at times $t=n/f_s$, $f_s= 10$~Hz. The 10~Hz signal was delayed by $D$ samples and compared to the original 10~MHz signal resampled at times $t-D/f_s$. The test results agreed with the calculated fractional error shown in Fig. \ref{fig:herror}.

\subsection{Further tests}
\label{Tests}

\begin{figure}[htb]
\begin{center}
\includegraphics[width=5.0 in]{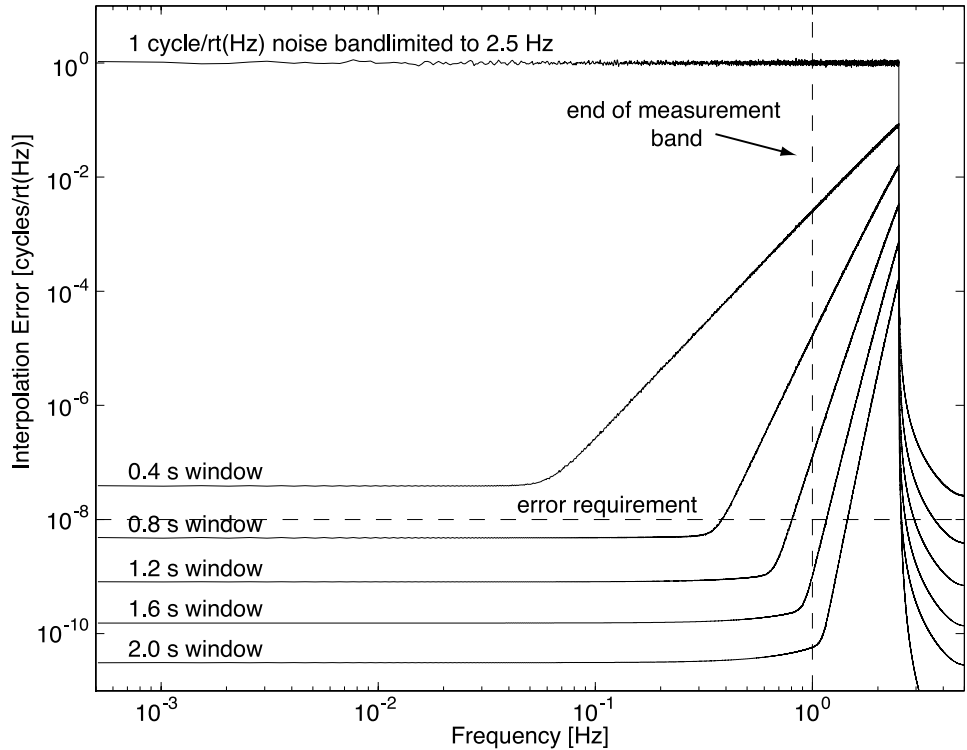}
\caption{Amplitude spectral density of interpolation error for Lagrange
  filters, shown with the spectral density
  of the initial 2.5~Hz-bandlimited noise. Spectral density is
  estimated by a triangle-windowed, averaged periodogram.
\label{fig:tditest}}
\end{center}
\end{figure}

We have characterized the error of fractional-delay filtering
with fixed delays.  In orbit, the delays will slowly vary due
to the changing arm lengths, and so we also tested Lagrange
filtering with varying delay.  For this
test, we generated a time series of white noise, bandlimited to
2.5~Hz and sampled at 10~Hz (oversampling factor of 2); we then used Lagrange filters of increasing order to interpolate the noise to the original
sampling times shifted by delays ranging linearly in time from $D=-0.5$ to $D=0$ (no delay), during a period of $5 \times
10^5$~seconds. This arrangement approximately simulates the slow
variation in the LISA armlengths (which determine the TDI-mandated
delays). Figure~\ref{fig:tditest} shows spectra of the interpolation
error, along with the spectrum of the original white noise. The required interpolation accuracy is achieved at all frequencies in the measurement band for $N \geq
16$ (window length of 1.6 s).

If the requirements become more stringent, for example due to
increased laser frequency noise, it is easy to improve the performance
to the desired level by increasing the length of the filter kernel.  By
setting the filter to $N=31$ (3 seconds in length at 10~Hz), $\varepsilon$ can be reduced to $1\times 10^{-15}$.

\section{Implications for LISA}

To illustrate the design simplifications enabled by high-precision
interpolation, consider the simplest TDI combination--the Michelson
combination $X(t)$~\cite{Tinto}:
\begin{equation}
\label{firstx}
X(t) = [s_{21}(t)-s_{31}(t)]-[s_{21}(t-2L_{3}/c)-s_{31}(t-2L_{2}/c)].
\end{equation}
where $s_{m1}$ is the phase measurement made at Spacecraft~1 of the
light received from Spacecraft $m$, $L_n$ is the length of the arm
opposite Spacecraft $n$, and we are assuming that the six LISA
lasers are phase locked~\cite{TintoPRD03}.  The
first two terms represent the optical phase of two arms of a Michelson
interferometer, and the second two terms represent the same quantity
with the specified delays.
\begin{figure}
\begin{center}
\includegraphics[scale=0.7]{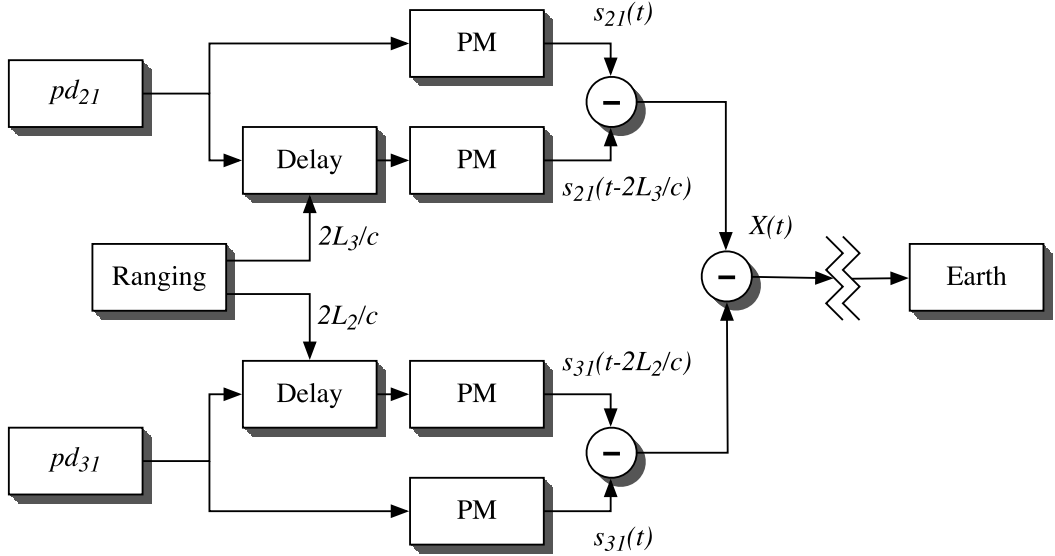}
\caption{Real-time TDI, in which the delays must be known at the time
  of measurement, and the TDI combinations are computed before
  transponding to Earth.  TDI combination $X(t)$ shown (Eq.~\ref{firstx}).}
\label{tdi-blocks}
\end{center}
\end{figure}

The implementation of TDI based on timed triggering \cite{TintoPRD03},
which we designate ``real-time TDI,'' calls for all four terms in
Eq.\ \eqref{firstx} to be explicitly measured, combined on-board, and sent to the
Earth. This is illustrated in Fig.~\ref{tdi-blocks} for the
measurement of $X(t)$ aboard Spacecraft~1.  The blocks labeled
$pd_{21}$ and $pd_{31}$ represent the radio-frequency beat signals
from the photodiode outputs, containing laser frequency noise
superimposed on the gravitational wave signal.  The Ranging block
represents the system that measures distances between spacecraft, and
computes the delays $2L_{3}/c$ and $2L_{2}/c$ required to assemble
$X(t)$.  The telemetry inputs to the ranging blocks (not shown)
contain ranging data measured on Spacecraft~1 and Spacecraft~2.  Phase
measurements are made by the phasemeter (PM) blocks.  In
Fig.~\ref{tdi-blocks}, the phase signals are delayed electronically by
the Delay blocks, which implement a variable delay as controlled by
the ranging system.  Equivalently, identical phase signals can be fed
to two PM blocks, and the timing of the phase measurements can be set
by adjustable triggers from the ranging system outputs.  The final
output $X(t)$ is free of laser frequency noise for fixed
armlengths. For time-dependent armlengths, we
expect~\cite{ShaddockPRD03} that velocity-correcting or ``second
generation'' TDI combinations will be required. They have roughly
double the measurements of length-correcting or ``first generation''
TDI combinations such as that shown in Fig.~\ref{tdi-blocks}.
\begin{figure}
\begin{center}
\includegraphics[scale=0.7]{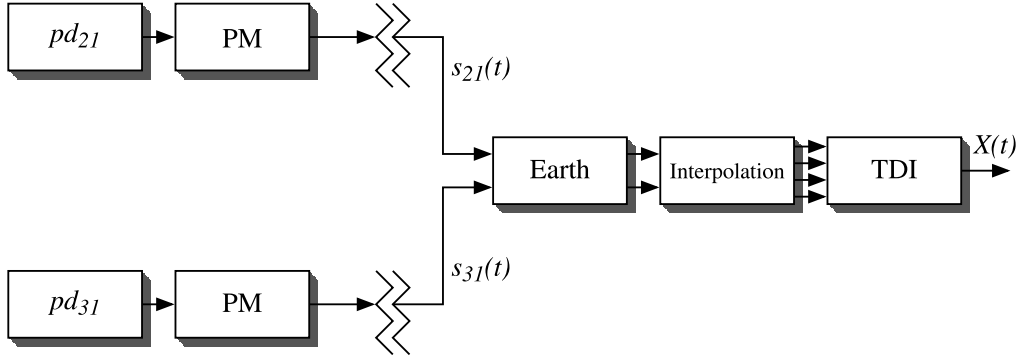}
\caption{Post-processed implementation of TDI combination $X(t)$.
\label{interp-blocks}}
\end{center}
\end{figure}

As we have demonstrated in this paper, the delayed phase measurements
required for TDI can alternatively be inferred by interpolation from an
equally-spaced sequence at a relatively low rate.  This implementation is referred to as post-processed TDI and is shown in Figure~\ref{interp-blocks}.  The elimination of the need for ranging
knowledge at the time of measurement simplifies the implementation, as
is evident comparing
Figs.~\ref{tdi-blocks}~and~\ref{interp-blocks}.  Ranging information
will still be needed as input to signal reconstruction, but it can be
transmitted to Earth independently of the phasemeter signals.
Alternatively, post-processed TDI can be implemented without explicit
ranging by determining the delays using autocorrelations
\cite{HellingsPRD01}, or by adjusting the delays in post-processing
for minimum sensitivity to laser frequency variations.  With
post-processed TDI it is no longer necessary to synchronize the clocks
on different spacecraft. Clock synchronization error can be corrected simply by time-shifting the data in post-processing.  The decoupling of ranging from phase measurements
and the removal of the need for clock synchronization allows
reduction--or possibly elimination--of inter-spacecraft
communications.

Post-processed TDI also allows complete flexibility in combining
phasemeter signals.  All the raw data are available for processing by
any TDI algorithm, including ones not developed until after the data
are in hand.  The delays can be adjusted to optimize the suppression
of laser frequency noise; by contrast, if there is an error in
triggering in real-time TDI, noise is irrevocably added.
Post-processed TDI simplifies the phase measurement hardware, allowing
all possible TDI combinations to be constructed from one constant-rate
phase measurement per photodetector.

A significant operating cost of LISA will be telemetry to Earth of
science data.  Real-time TDI requires one data stream per TDI
combination.  Post-processed TDI requires one data stream per phasemeter;
more precisely, per phasemeter that does not have its output held
fixed by a high-gain control system.  We expect that post-processed
TDI will require fewer telemetry signals than real-time TDI, but this
depends on details of hardware design and on data requirements that
are currently under consideration.

Other factors influencing overall telemetry costs are the data rate
and number of bits per datum.  The ultimate LISA data will have a
signal bandwidth of 1~Hz and a dynamic range large enough to encompass
both the largest expected gravitational wave signal (or perhaps the
largest instrumental effect) and shot-noise limited sensitivity.  Each output
of the real-time TDI signal chain ($X(t)$ for example) will have
essentially eliminated laser frequency noise before data are
transponded to Earth, reducing the dynamic range requirement.  A
nominal datum size is 20~bits per sample.  The data rate for each TDI
combination is 2~samples/second, in keeping with the 1~Hz requirement and
the Nyquist limit.

In comparison, post-processed TDI must transpond large laser frequency
noise superimposed on the small gravitational wave signal.  Larger dynamic range is required, perhaps 30~bits per sample.  The sample rate for post-processed TDI is likely to be greater than 2~samples/second for two reasons. Firstly a larger sampling frequency may be needed to provide the more stringent anti-aliasing filtering needed when laser frequency noise is present. Secondly an oversampling factor may be needed for the interpolation procedure.  The minimum sample rate imposed by post-processed TDI is still under study; our initial estimate is 10~samples/second (oversampling factor of 5) or less.

Although the interpolation algorithms presented here may not be
optimal, they demonstrate the feasibility of post-processed TDI.  They
serve as a proof of principle, and provide guidance for the design of
LISA with significant simplification in several respects over
real-time TDI.

\section{Acknowledgments}

We thank John Armstrong for many useful discussions and for help with tests of the interpolation procedure. M.V.\ was supported by the LISA
Mission Science Office at JPL. This research was
performed at the Jet Propulsion Laboratory, California Institute of
Technology, under contract with the National Aeronautics and Space
Administration.

\end{document}